\documentclass[namedreferences]{kluwer}
\usepackage{QED}
\usepackage[PostScript=dvips]{diagrams}
\usepackage{euscript}

\newtheorem{theorem}{Theorem}[section]
\newtheorem{lemma}[theorem]{Lemma}
\newtheorem{proposition}[theorem]{Proposition}

\newtheorem{question}[theorem]{Question}

\newcommand{\Imp}{\; \Rightarrow \;}

\newcommand{\IFF}{\;\; \Longleftrightarrow \;\;}
\newcommand{\Iff}{\; \Leftrightarrow \;}
\newcommand{\id}{\mathsf{id}}
\newcommand{\Set}{\mathbf{Set}}
\newcommand{\Chu}{\mathbf{Chu}}
\newcommand{\rarr}{\rightarrow}
\newcommand{\pow}{\mathcal{P}}
\newcommand{\CC}{\EuScript{C}}
\newcommand{\DD}{\EuScript{D}}
\newcommand{\HH}{\mathcal{H}}
\newcommand{\KK}{\mathcal{K}}
\newcommand{\LL}{\mathsf{L}}
\newcommand{\PP}{\mathsf{P}}

\newcommand{\Ho}{\HH_{\circ}}
\newcommand{\So}{S_{\circ}}
\newcommand{\Uo}{U_{\circ}}
\newcommand{\Ko}{\KK_{\circ}}
\newcommand{\ip}[2]{\langle #1 \mid #2 \rangle}
\newcommand{\norm}[1]{\| #1  \|}
\newcommand{\ray}[1]{\bar{#1}}
\newcommand{\eH}{e_{\HH}}
\newcommand{\eK}{e_{\KK}}
\newcommand{\beH}{\bar{e}_{\HH}}
\newcommand{\beK}{\bar{e}_{\KK}}
\newcommand{\Complex}{\mathbb{C}}
\newcommand{\Real}{\mathbb{R}}
\newcommand{\FF}{\mathbb{F}}
\newcommand{\GG}{\mathbb{G}}
\newcommand{\fL}{f^{\rarr}}
\newcommand{\bC}{\mathbf{bChu}}
\newcommand{\bmC}{\mathbf{bmChu}}
\newcommand{\sC}{\mathbf{sChu}}
\newcommand{\eC}{\mathbf{eChu}}
\newcommand{\emC}{\mathbf{emChu}}

\newcommand{\ie}{\textit{i.e.}~}
\newcommand{\Image}{\mathrm{Im \,}}
\newcommand{\orth}{\, \bot \,}

\newcommand{\SymmH}{\mathbf{SymmH}}
\newcommand{\PSymm}{\mathbf{\PP SymmH}}
\newcommand{\UG}{\mathsf{U}(1)}
\newcommand{\nn}{\mathbf{n}}
\newcommand{\Two}{\mathbf{2}}
\newcommand{\Three}{\mathbf{3}}
\newcommand{\Fv}{F_{v}}
\newcommand{\zv}{\mathbf{0}}

\newarrow{inc}C--->
\newarrow{mon}>--->
\newarrow{epi}----{>>}
\newarrow{rep}>---{>>}

\begin{document}
\begin{article}
\begin{opening}

\title{Big Toy Models}
\subtitle{Representing Physical Systems As Chu Spaces}
\author{Samson \surname{Abramsky}}
\institute{Oxford University Computing Laboratory}

\begin{abstract}
We pursue a model-oriented rather than axiomatic approach to the foundations of Quantum Mechanics, with the idea that new models can often suggest new axioms. This approach has often been fruitful in Logic and Theoretical Computer Science. Rather than seeking to construct a simplified toy model, we aim for a `big toy model', in which both quantum and classical systems can be faithfully represented --- as well as, possibly, more exotic kinds of systems.
To this end, we show how Chu spaces can be used to represent physical systems of various kinds. In particular, we show how quantum systems can be represented as Chu spaces over the unit interval in such a way that the Chu morphisms correspond exactly to the physically meaningful symmetries of the systems --- the unitaries and antiunitaries. In this way we obtain a full and faithful functor from the groupoid of Hilbert spaces and their symmetries to Chu spaces. We also consider whether it is possible to use a finite value set rather than the unit interval; we show that three values suffice, while the two standard possibilistic reductions to two values both fail to preserve fullness.
\end{abstract}

\end{opening}

\section{Introduction}

\paragraph{Models vs. Axioms}
The main method pursued in the foundations of quantum mechanics has been \emph{axiomatic}; one seeks conceptually primitive and clearly motivated axioms, shows that quantum systems satisfy these axioms, and then, often, aims for a \emph{representation theorem} showing that the axioms essentially determine the ``standard model'' of Quantum Mechanics. Or one may admit non-standard interpretations, and seek to locate Quantum Mechanics in a larger ``space'' of theories.

There is an alternative and complementary approach, which has been less explored in the foundations of Quantum Mechanics, although it has proved very fruitful in mathematics, logic and theoretical computer science. Namely, one looks for conceptually natural \emph{constructions of models}. Often a new model construction can suggest new axioms, articulated in terms of new forms of structure. There are many examples of this phenomenon, sheaves and topos theory being one case in point \cite{MacMoe92}, and domain-theoretic models of the $\lambda$-calculus another \cite{Scott70}.

A  successful recent example of gaining insight by model construction is the well-known paper by Rob Spekkens on a toy model for Quantum Mechanics \cite{Spekk07}, which has led to novel ideas on the connections between phase groups and non-locality \cite{RR-09-04}.

\paragraph{Big Toy Models}
We shall also, in a sense, be concerned with ``toy models'' in the present paper; with building models which exhibit ``quantum-like'' features without necessarily exactly corresponding to the standard formalism of Quantum Mechanics. Indeed, the more different the model construction can be to the usual formalism, while still reproducing many quantum-like features, the more interesting it will be from this perspective.
However, there will be an important difference between the kind of model we shall study, and the usual idea of a ``toy model''. Usually, a toy model will be a small, simplified gadget, which gives a picture of Quantum Mechanics in some ``collapsed'' form, with much detail thrown away. By contrast, we are aiming for a \emph{big} toy model, in which \emph{both quantum and classical systems can be faithfully represented} --- as well as, possibly, many more exotic kinds of systems.

\paragraph{Results}
More precisely, we shall see how the simple, discrete notions of Chu spaces suffice to determine the appropriate notions of state equivalence, and to pick out the physically significant symmetries on Hilbert space in a very striking fashion. This leads to a full and faithful representation of the category of quantum systems, with the groupoid structure of their physical symmetries,  in the category of Chu spaces valued in the unit interval. The arguments here make use of Wigner's theorem and the dualities of projective geometry, in the modern form developed by Faure and Fr\"olicher \cite{FauFro00,SvS07}. The surprising point is that unitarity/anitunitarity is essentially \emph{forced} by the mere requirement of being a Chu morphism. This even extends to surjectivity, which here is \emph{derived} rather than assumed.

We also consider the question of whether we can obtain a natural representation of this form in Chu spaces over a finite value set. We show that this can be done with just three values. By contrast, the two standard possibilistic reductions to two values both \emph{fail to preserve fullness}. 

The use of Chu spaces for representing physical systems as initiated in this paper seems quite promising; a number of further topics immediately suggest themselves, including mixed states, universal models, the representation of convex theories, linear types, and local logics for quantum systems.

The plan of the remainder of the paper is as follows. In Section~2, we shall provide a brief overview of Chu spaces. Section~3 contains the main technical results, leading to a full and faithful representation of quantum systems and their symmetries as Chu spaces and morphisms of Chu spaces. Section~4 presents the results on finite value sets. Finally, Section~5 contains a discussion of conceptual and methodological issues.

\section{Chu Spaces}
We shall assume that the reader is familiar with a few basic notions of category theory.\endnote{The charming introductory text \cite{Pie91} should be more than sufficient.} The bare definitions of category and functor will suffice for the most part.

Chu spaces are a special case of a construction which originally appeared in \cite{Chu79},  written by Po-Hsiang Chu as an appendix to Michael Barr's monograph on $*$-autonomous categories \cite{Barr79}.

Interest in $*$-autonomous categories increased with the advent of Linear Logic \cite{Gir87}, since $*$-autonomous categories provide models for Classical Multiplicative Linear Logic (and with additional assumptions, for the whole of Classical Linear Logic) \cite{See89}.
The Chu construction applied to the category $\Set$ of sets and functions was independently introduced (under the name of `games') by Yves Lafont and Thomas Streicher \cite{LS91}, and subsequently (under the name of \emph{Chu spaces}) formed the subject of a series of papers by Vaughan Pratt and his collaborators, e.g. \cite{DBLP:conf/lics/DevarajanHPP99,DBLP:conf/lics/Pratt95,DBLP:journals/apal/Pratt99}. Recent papers on Chu spaces include \cite{DBLP:conf/calco/DrosteZ07,DBLP:conf/calco/PalmigianoV07}.

Chu spaces have several interesting aspects:
\begin{itemize}
\item They have a rich type structure, and in particular form models of Linear Logic.
\item They have a rich representation theory; many concrete categories of interest  can be fully embedded into Chu spaces.
\item There is a natural notion of `local logic' on Chu spaces \cite{Barwise97}, and an interesting characterization of information transfer across Chu morphisms \cite{DBLP:journals/igpl/Benthem00}.
\end{itemize}
Applications of Chu spaces have been proposed in a number of areas, including concurrency \cite{DBLP:journals/mscs/Pratt03},
hardware verification \cite{DBLP:conf/cdes/Ivanov08}, game theory \cite{RePEc:usi:wpaper:417}  and fuzzy systems \cite{Pap00,DBLP:journals/jaciii/NguyenNWK01}. Mathematical studies concerning the general Chu construction include \cite{Barr98,DBLP:journals/acs/GiuliT07}.

We briefly review the basic definitions.

Fix a set $K$. A Chu space over $K$ is a structure $(X, A, e)$, where $X$ is a set of `points' or `objects', $A$ is a set of `attributes', and $e : X \times A \rarr K$ is an evaluation function.

A morphism of Chu spaces
\[ f : (X, A, e) \rarr (X', A', e') \]
is a pair of functions
\[ f = (f_{*} : X \rarr X', f^{*} : A' \rarr A) \]
such that, for all $x \in X$ and $a' \in A'$:
\[ e(x, f^{*}(a')) = e'(f_{*}(x), a') . \]
Chu morphisms compose componentwise: if $f : (X_{1}, A_{1}, e_{1}) \rarr (X_{2}, A_{2}, e_{2})$ and $g : (X_{2}, A_{2}, e_{2}) \rarr (X_{3}, A_{3}, e_{3})$, then
\[ (g \circ f)_{*} = g_{*} \circ f_{*}, \qquad (g \circ f)^{*} = f^{*} \circ g^{*} . \]
Chu spaces over $K$ and their morphisms form a category $\Chu_{K}$.

Given a Chu space $C = (X, A, e)$, we say that $C$ is:
\begin{itemize}
\item \emph{extensional} if for all $a_{1}, a_{2} \in A$:
\[ [\forall x \in X. \, e(x, a_{1}) = e(x, a_{2})] \Imp a_{1} = a_{2} \]
\item \emph{separated} if for all $x_{1}, x_{2} \in X$:
\[ [\forall a \in A. \, e(x_{1}, a) = e(x_{2}, a)] \Imp x_{1} = x_{2} \]
\item \emph{biextensional} if it is extensional and separated.
\end{itemize}
We define a relation on $X$ by:
\[ x_{1} \sim x_{2}  \IFF \forall a \in A. \, e(x_{1}, a) = e(x_{2}, a) . \]
This is evidently an equivalence relation: $C$ is separated exactly when this relation is the identity. There is a Chu morphism
\[ (q, \id_{A}) : (X, A, e) \rarr (X/{\sim}, A, e') \]
where $e'([x], a) = e(x, a)$ and $q : X \rarr X/{\sim}$ is the quotient map. The space $(X/{\sim}, A, e')$ is separated; if $(X, A, e)$ is extensional, it is biextensional.

\begin{proposition}
\label{simpresprop}
If $f : (X, A, e) \rarr (X', A', e')$ is a Chu morphism, then $f_*$ preserves $\sim$. That is, for all $x_1, x_2 \in X$,
\[ x_1 \sim x_2 \Imp f_*(x_1) \sim f_*(x_2) . \]
\end{proposition}
\begin{proof}
For any $a' \in A'$:
\[ e'(f_*(x_1), a') = e(x_1, f^*(a')) = e(x_2, f^*(a')) = e'(f_*(x_2), a') . \]
\end{proof}

We shall write $\eC_K$, $\sC_K$ and $\bC_K$ for the full subcategories of $\Chu_K$ determined by the extensional, separated and biextensional Chu spaces.

We shall mainly work with extensional and biextensional Chu spaces. Obviously $\bC_K$ is a full sub-category of $\eC_K$.
\begin{proposition}
\label{bcollprop}
The inclusion $\bC_K \rinc \eC_K$ has a left adjoint.
\end{proposition}
\begin{proof}
The unit of the adjunction is the Chu morphism
\[ (q, \id_{A}) : (X, A, e) \rarr (X/{\sim}, A, e') \]
we have already described, while Proposition~\ref{simpresprop} guarantees that given a Chu morphism
\[ f : (X, A, e) \rarr (Y, B, r) \]
to a biextensional Chu space, we can factor it through the quotient space $(X/{\sim}, A, e')$.

The functor $Q : \eC_K \rarr \bC_K$ provided by this adjunction sends morphisms 
\[ (f_*, f^*) : (X, A, e_1) \rarr (X', A', e_2) \]
to
\[ (f_*/{\sim}, f^*) : (X/{\sim}, A, e_1') \rarr (X'/{\sim}, A', e_2') \]
where $f_*/{\sim}([x]) = [f_*(x)]$.
\end{proof}
We refer to the functor $Q$ as the \emph{biextensional collapse}.

We can define an equivalence relation on the Chu morphisms in each hom-set in $\eC_K$ by:
\[ f \sim g \IFF \forall x. \, f_*(x) \sim g_*(x) . \]
Then $Qf = Qg \Iff f \sim g$.

\paragraph{Representations}
Recall that a functor $F : \CC \rarr \DD$ is \emph{faithful} if for each pair of objects $A$, $B$ of $\CC$, the induced map $F_{AB} : \CC(A, B) \rarr \DD(FA, FB)$ is injective; it is \emph{full} if each $F_{AB}$ is surjective; and it is an \emph{embedding} if $F$ is faithful and injective on objects. We refer to a full and faithful functor as a \emph{representation}, and to a full embedding as a \emph{strict representation}. Note that if $F$ is a representation, it can only identify isomorphic objects. If $F$ is a representation, then $\CC$ is equivalent to a full sub-category of $\DD$, while if $F$ is a strict representation,  then $\CC$ is isomorphic to a full sub-category of $\DD$.

As a first example of the representational capacity of Chu spaces, suppose that $\{ 0, 1\} \subseteq K$. For any set $X$, define the following Chu space on $K$: $(X, \pow X, e_{X})$, where:
\[  e_{X}(x, S) = \left\{ \begin{array}{lr}
1, & x \in S \\
0 & \mbox{otherwise}
\end{array}
\right.
\]
Given a function $f : X \rarr Y$, we send it to the Chu space morphism
\[ (f, f^{-1}) : (X, \pow X, e_{X}) \rarr (Y, \pow Y, e_{Y}) . \]
It is easy to see that this defines a full embedding of $\Set$ into $\Chu_{K}$.

\section{Representation of Quantum Systems}

Our point of view in modelling physical systems as Chu spaces will be as follows. We take a system to be specified by its set of \emph{states} $S$, and the set of \emph{questions} $Q$ which can be `asked' of the system. We shall consider only `yes/no' questions; however, the result of asking a question in a given state will in general be \emph{probabilistic}. This will be represented by an evaluation function
\[ e : S \times Q \rarr [0, 1] \]
where $e(s, q)$ is the probability that the question $q$ will receive the answer `yes' when the system is in state $s$. This is essentially the point of view taken by Mackey in his classic pioneering work on the foundations of Quantum Mechanics \cite{Mack63}. Note that, following Mackey, we prefer the term `question' to `property', since in the case of Quantum Mechanics we cannot think in terms of static properties which are determinately possessed by a given state; questions imply a dynamic \emph{act} of asking. 

It is standard in the foundational literature on quantum mechanics to focus on yes/no questions. However, the usual approaches to quantum logic avoid the direct introduction of probabilities. We shall return to the issue of whether it is necessary to take probabilities as our value set in Section~\ref{valsect}.

We can take the category $\Set$ itself as a crude version of discrete deterministic classical systems, with arbitrary irreversible transformations allowed. We now consider the quantum case, in the pure state formulation. Mixed states will be considered in a sequel to the present paper.

Let $\HH$ be a complex Hilbert space.\endnote{A useful reference for the mathematical background is \cite{Jor69}.}
We define the following Chu space over $[0, 1]$:
\[ (\Ho, \LL(\HH), e_{\HH}) \]
where:
\begin{itemize}
\item $\Ho = \HH \setminus \{ \mathbf{0} \}$, the set of non-zero vectors. We shall regard all such vectors, not necessarily normalized, as representations of states of the system. Note that the zero vector is \emph{not} a legitimate state; its r\^ole in Quantum Mechanics proper (as opposed to linear-algebraic calculations) is largely as an `error element' when operations cannot legitimately be performed.
\item $\LL(\HH)$ is the lattice of closed subspaces of $\HH$. This is the standard notion of yes/no questions in Quantum Mechanics. The observable corresponding to the subspace $S$ is the self-adjoint operator whose spectral decomposition is $S \oplus S^{\bot} \cong \HH$. To each subspace $S$ there corresponds the projector $P_S$.
\item The evaluation $\eH$ is the fundamental formula or `statistical algorithm' \cite{Red87} giving the basic predictive content of Quantum Mechanics:
\[ \eH(\psi, S) = \frac{\ip{\psi}{P_S \psi}}{\ip{\psi}{\psi}} = \frac{\ip{P_S \psi}{P_S \psi}}{\ip{\psi}{\psi}}  = \frac{\norm{P_S \psi}^2}{\norm{\psi}^2} .  \]
Note that $\eH(\psi, S) = \eH(\frac{\psi}{\norm{\psi}}, S)$, so this is equivalent to working with normalized vectors only.
\end{itemize}
We have thus directly transcribed the basic ingredients of the Dirac/von Neumann-style formulation of Quantum Mechanics \cite{Dirac,vN} into the definition of the Chu space corresponding to a given Hilbert space.

\subsection{Characterizing Chu Morphisms on Quantum Chu Spaces}

Recall firstly the following explicit expression for the projection of a vector $\psi$ on a subspace $S$.
Let $\{ e_{i} \}$ be an orthonormal basis for $S$. Then
\[ P_{S} \psi  = \sum_{i} \ip{\psi}{e_{i}} e_{i} . \]
It follows  that $\psi \orth S$ if and only if $P_{S} \psi = \mathbf{0}$.

We begin with a basic fact which we record explicitly.

\begin{lemma}
\label{memblemm}
For $\psi \in \Ho$ and $S \in \LL(\HH)$:
\[ \psi \in S \IFF \eH(\psi, S) = 1 . \]
\end{lemma}
\begin{proof}
Firstly, if $\psi \in S$, then $P_S(\psi) = \psi$, so $\eH(\psi, S) = 1$.

Next, we recall that $P_{S^{\bot}} = I - P_S$.
Hence 
\[ \begin{array}{rcl}
\eH(\psi, S^{\bot})  & = &  \frac{1}{\ip{\psi}{\psi}} \ip{\psi - P_S \psi}{\psi - P_S \psi} \\
 & = & \frac{1}{\ip{\psi}{\psi}} (\ip{\psi}{\psi} - \ip{\psi}{P_S \psi} - \ip{P_S \psi}{\psi} + \ip{P_S \psi}{P_S \psi}) \\
 & = &  \frac{1}{\ip{\psi}{\psi}} (\ip{\psi}{\psi} - \ip{P_S \psi}{P_S \psi}).
 \end{array}
\]
Hence
\[ \begin{array}{rcl}
\eH(\psi, S) + \eH(\psi, S^{\bot}) & = & \frac{1}{\ip{\psi}{\psi}} (\ip{P_S \psi}{P_S \psi} + \ip{\psi}{\psi} - \ip{P_S \psi}{P_S \psi}) \\
& = & \frac{1}{\ip{\psi}{\psi}} \ip{\psi}{\psi} = 1 . 
\end{array}
\]
So if $\psi \not\in S$, it suffices to show that $\eH(\psi, S^{\bot}) > 0$. In this case, $\psi = \theta + \chi$, where $\theta \in S$ and $\chi \in S^{\bot} \setminus \{ \mathbf{0} \}$; so $P_{S^{\bot}}(\theta) = \mathbf{0}$ and $P_{S^{\bot}}(\chi) = \chi$. Then
\[ \begin{array}{rcl}
\eH(\psi, S^{\bot}) & = & \frac{1}{\ip{\psi}{\psi}} \ip{P_{S^{\bot}}(\theta) + P_{S^{\bot}}(\chi)}{P_{S^{\bot}}(\theta) + P_{S^{\bot}}(\chi)} \\
& = & \frac{1}{\ip{\psi}{\psi}} \ip{\chi}{\chi}  > 0 . 
\end{array}
\]
\end{proof}

\begin{proposition}
\label{extprop}
The Chu space $(\Ho, \LL(\HH), e_{\HH})$ is extensional but not separated. The equivalence classes of the relation $\sim$ on states are exactly the \emph{rays} of $\HH$. That is:
\[ \phi \sim \psi \IFF \exists \lambda \in \Complex . \, \phi = \lambda \psi . \]
\end{proposition}
\begin{proof}
Extensionality follows directly from Lemma~\ref{memblemm}, since if two subspaces have the same evaluations on all states, they have the same elements.

We have
\[ \eH(\lambda \psi, S) = \frac{|\lambda|^2}{|\lambda|^2} \frac{\ip{P_S \psi}{P_S \psi}}{\ip{\psi}{\psi}} = \eH(\psi, S)  \]
so $\phi = \lambda \psi \Imp \phi \sim \psi$. For the converse, let $S$ be the one-dimensional subspace (ray) spanned by $\psi$, and suppose that $\phi \not\in S$. By Lemma~\ref{memblemm}, $\eH(\psi, S) = 1$, while $\eH(\phi, S) \neq 1$. Hence $\phi \not\sim \psi$.
\end{proof}

Thus we have recovered the standard notion of pure states as the rays of the Hilbert space from the general notion of state equivalence in Chu spaces.

We shall now use some notions and results from projective geometry. We shall use the very nice Handbook article \cite{SvS07} as a convenient reference.

Given a vector $\psi \in \Ho$, we write $\ray{\psi} = \{ \lambda\psi \mid \lambda \in \Complex \}$ for the ray which it generates. The rays are the \emph{atoms} in the lattice $\LL(\HH)$.

We write $\PP(\HH)$ for the set of rays of $\HH$. By virtue of Proposition~\ref{extprop}, we can write the biextensional collapse of $(\Ho, \LL(\HH), e_{\HH})$ given by Proposition~\ref{bcollprop} as
\[ (\PP(\HH), \LL(\HH), \beH) \]
where $\beH(\ray{\psi}, S) = \eH(\psi, S)$.

We restate Lemma~\ref{memblemm} for the biextensional case.

\begin{lemma}
\label{inclemm}
For $\psi \in \Ho$ and $S \in \LL(\HH)$:
\[ \beH(\ray{\psi}, S) = 1 \IFF \ray{\psi} \subseteq S . \]
\end{lemma}
\begin{proof}
Since $S$ is a subspace, $\ray{\psi} \subseteq S$ iff $\psi \in S$, and the result follows from Lemma~\ref{memblemm}.
\end{proof}

We now turn to the  issue of characterizing the Chu morphisms between these biextensional Chu representations of Hilbert spaces.
This will lead to our first representation theorem.

To fix notation, suppose we have Hilbert spaces $\HH$ and $\KK$, and a Chu morphism
\[ (f_*, f^*) : (\PP(\HH), \LL(\HH), \beH) \rarr (\PP(\KK), \LL(\KK), \beK) . \]

\begin{proposition}
\label{padjprop}
For $\psi \in \Ho$ and $S \in \LL(\KK)$:
\[ \ray{\psi} \subseteq f^*(S) \IFF f_*(\ray{\psi}) \subseteq S . \]
\end{proposition}
\begin{proof}
By Lemma~\ref{inclemm}:
\[ \ray{\psi} \subseteq f^*(S) \Iff \beH(\ray{\psi}, f^*(S)) = 1 \Iff \beK(f_*(\ray{\psi}), S) = 1 \Iff f_*(\ray{\psi}) \subseteq S . \]
\end{proof}

Note that $\PP(\HH) \subseteq \LL(\HH)$.

\begin{proposition}
\label{invinjprop}
The following are equivalent:
\begin{itemize}
\item $f_*$ is injective
\item The following diagram commutes:
\end{itemize}
\begin{equation}
\label{incdia}
\begin{diagram}
\PP(\HH) & \rTo^{f_*} & \PP(\KK) \\
\dinc & & \dinc \\
\LL(\HH) & \lTo_{f^*} & \LL(\KK)
\end{diagram}
\end{equation}
That is, for all $\psi \in \Ho$:
\[ \ray{\psi} = f^*(f_*(\ray{\psi})). \]
\end{proposition}
\begin{proof}
Clearly, (\ref{incdia}) implies that $f_{*}$ is injective. For the converse,
Proposition~\ref{padjprop} implies that $\ray{\psi} \subseteq f^*(f_*(\ray{\psi}))$. Now suppose that $\ray{\phi} \subseteq f^*(f_*(\ray{\psi}))$. Applying Proposition~\ref{padjprop} again, this implies that $f_*(\ray{\phi}) \subseteq f_*(\ray{\psi})$. Since  $f_*(\ray{\phi})$ and $f_*(\ray{\psi})$ are atoms, this implies that $f_*(\ray{\phi}) = f_*(\ray{\psi})$, which since $f_*$ is injective implies that $\ray{\phi} = \ray{\psi}$. Thus the only atom below $f^*(f_*(\ray{\psi}))$ is $\ray{\psi}$. Since $\LL(\HH)$ is \emph{atomistic} \cite{SvS07}, this implies that $f^*(f_*(\ray{\psi})) \subseteq \ray{\psi}$.
\end{proof}

We state another important basic property of the evaluation.
\begin{lemma}
\label{orthlemm}
For any $\phi, \psi \in \Ho$:
\[ \beH(\ray{\phi}, \ray{\psi}) = 0 \IFF \phi \orth \psi . \]
\end{lemma}
\begin{proof}
\[ \beH(\ray{\phi}, \ray{\psi}) = 0 \Iff \ip{P_{\ray{\psi}}(\phi)}{P_{\ray{\psi}}(\phi)} = 0 \Iff P_{\ray{\psi}}(\phi) = \mathbf{0} \Iff \phi \orth \psi . \]
\end{proof}

\begin{proposition}
\label{orthpresprop}
If $f_*$ is injective, it \emph{preserves and reflects orthogonality}. That is, for all $\phi, \psi \in \Ho$:
\[ \phi \orth \psi \IFF f_*(\ray{\phi}) \orth f_*(\ray{\psi}) . \]
\end{proposition}
\begin{proof}
\[ \begin{array}{lclr}
\phi \orth \psi & \Longleftrightarrow & \beH(\ray{\phi}, \ray{\psi}) = 0 & \mbox{Lemma \ref{orthlemm}} \\
& \Longleftrightarrow & \beH(\ray{\phi}, f^*(f_*(\ray{\psi}))) = 0 & \mbox{Proposition~\ref{invinjprop}} \\
& \Longleftrightarrow & \beK(f_*(\ray{\phi}), f_*(\ray{\psi})) = 0 &  \\
& \Longleftrightarrow &  f_*(\ray{\phi}) \orth f_*(\ray{\psi}) & \mbox{Lemma \ref{orthlemm}}
\end{array}
\]
\end{proof}

We define a map $\fL : \LL(\HH) \rarr \LL(\KK)$:
\[ \fL(S) = \bigvee \{ f_*(\ray{\psi}) \mid \psi \in \So \} \]
where $\So = S \setminus \{ \mathbf{0} \}$.

\begin{lemma}
The map $\fL$ is left adjoint to $f^*$.
\end{lemma}
\begin{proof}
We must show that, for all $S \in \LL(\HH)$ and $T \in \LL(\KK)$:
\[ \fL(S) \subseteq T \IFF S \subseteq f^*(T) . \]
Using Proposition~\ref{padjprop}, we have:
\[ \begin{array}{lcl}
\fL(S) \subseteq T & \Longleftrightarrow & \forall \psi \in \So . \, f_*(\ray{\psi}) \subseteq T \\
& \Longleftrightarrow & \forall \psi \in \So . \, \ray{\psi} \subseteq f^*(T) \\
& \Longleftrightarrow & S \subseteq f^*(T) .
\end{array}
\]
\end{proof}

We can now extend the diagram ~(\ref{incdia}):
\begin{equation}
\label{incdia2}
\begin{diagram}
\PP(\HH) & \rTo^{f_*} & \PP(\KK) \\
\dinc & & \dinc \\
\LL(\HH) & \pile{\rTo^{\fL}\\ \bot \\ \lTo_{f^*}}  & \LL(\KK)
\end{diagram}
\end{equation}

By construction, $\fL$ extends $f_*$: this says that $\fL$ preserves atoms.
Since $\fL$ is a left adjoint, it preserves sups. Hence $\fL$ and $f_*$ are paired under the duality of projective lattices and projective geometries, for which see Theorem~16 of \cite{SvS07}. In particular, 
we have the following.

\begin{proposition}
\label{projmapprop}
$f_*$ is a \emph{total map of projective geometries} \cite{SvS07}.
\end{proposition}

It follows that we can apply  \emph{Wigner's Theorem}, in the form given as Theorem~4.1 in \cite{Fau02}. In order to state this, we need some additional notions.

Let $V_1$ be a vector space over the field $\FF$ and $V_2$ a vector space over the field $\GG$. A \emph{semilinear map} from $V_1$ to $V_2$ is a pair $(f, \alpha)$ where $\alpha : \FF \rarr \GG$ is a field homomorphism, and $f : V_1 \rarr V_2$ is an additive map such that, for all $\lambda \in \FF$ and $v \in V_1$:
\[ f(\lambda v) = \alpha(\lambda) f(v) . \]
Note that semilinear maps compose: if $(f, \alpha) : V_1 \rarr V_2$ and $(g, \beta) : V_2 \rarr V_3$, then $(g \circ f, \beta \circ \alpha) : V_1 \rarr V_2$ is a semilinear map.

This notion is usually defined in greater generality, for division rings, but we are only concerned with Hilbert spaces over the complex numbers.

Given a semilinear map $g : V_1 \rarr V_2$, we define $\PP g : \PP V_1 \rarr \PP V_2$ by
\[ \PP(g)(\ray{\psi}) = \overline{g(\psi)} . \]

We can now state Wigner's Theorem in the form we shall use it. 
\begin{theorem}
\label{Wigth}
Let $f : \PP(\HH) \rarr \PP(\KK)$ be a total map of projective geometries, where $\dim \HH > 2$. If $f$ preserves orthogonality, meaning that
\[ \ray{\phi} \orth \ray{\psi} \Imp f(\ray{\phi}) \orth f(\ray{\psi}) \]
then there is a semilinear map $g : \HH \rarr \KK$ such that $\PP(g) = f$, and 
\[ \ip{g(\phi)}{g(\psi)} = \sigma(\ip{\phi}{\psi}) , \]
where $\sigma$ is the homomorphism associated with $g$. Moreover, this homomorphism is either the identity or complex conjugation, so $g$ is either linear or antilinear. The map $g$ is unique up to a \emph{phase}, \ie a scalar of modulus 1.
\end{theorem}
The final statement follows from the Second Fundamental Theorem of Projective Geometry, Theorem~3.1 in \cite{Fau02} or Theorem~46 in \cite{SvS07}.

Note that  in our case, taking $f_* = f$, $\PP g$ is just the action of the biextensional collapse functor on Chu morphisms.

Note that a total map of projective geometries must necessarily come from an \emph{injective} map $g$ on the underlying vector spaces, since $\PP(g)$ maps rays to rays, and hence $g$ must have trivial kernel. For this reason, partial maps of projective geometries are considered in the Faure-Fr\"olicher approach \cite{FauFro00,SvS07}.
However, we are simply following the `logic' of Chu space morphisms here.

\begin{proposition}
\label{surjprop}
Let $g : \HH \rarr \KK$ be a semilinear morphism such that $\PP(g) = f_*$ where $f$ is a Chu space morphism, and $\dim (\HH) > 0$. Suppose that the endomorphism $\sigma : \Complex \rarr \Complex$ associated with $g$ is surjective, and hence an automorphism. Then $g$ is surjective.
\end{proposition}
\begin{proof}
We write $\Image g$ for the set-theoretic direct image of $g$, which is a linear subspace of $\KK$, since $\sigma$ is an automorphism. In particular, $g$ carries rays to rays, since $\lambda g(\phi) = g(\sigma^{-1}(\lambda) \phi)$.

We claim that for any vector $\psi \in \Ko$ which is not in the image of $g$, $\psi \orth \Image g$.
Given such a $\psi$, for any $\phi \in \Ho$ it is not the case that $f_*(\ray{\phi}) \subseteq \ray{\psi}$; for otherwise, for some $\lambda$, $g(\phi) = \lambda\psi$, and hence $g(\sigma^{-1}(\lambda^{-1})\phi) = \psi$. Then by Proposition~\ref{padjprop}, $f^*(\ray{\psi}) = \{ \mathbf{0} \}$. It follows that for all $\phi \in \Ho$, 
\[ \beK(f_*(\ray{\phi}), \ray{\psi}) = \beH (\ray{\phi},  \{ \mathbf{0} \}) = 0, \]
and hence by Lemma~\ref{orthlemm}  that $\psi \orth \Image g$. 

Now suppose for a contradiction that such a $\psi$ exists.
Consider the vector $\psi + \chi$ where $\chi$ is a non-zero vector in $\Image g$, which must exist since $g$ is injective and $\HH$ has positive dimension. This vector is not in $\Image g$, nor is it orthogonal to $\Image g$, since e.g. $\ip{\psi + \chi}{\chi} = \ip{\chi}{\chi} \neq 0$. This  yields the required contradiction.
\end{proof}

We can now put the pieces together to obtain the main result of this section.
We say that a map $U : \HH \rarr \KK$ is \emph{semiunitary} if it is either unitary or antiunitary; that is, if it is a bijective map satisfying
\[ U(\phi + \psi) = U\phi + U\psi, \qquad U(\lambda\phi) = \sigma(\lambda)U\phi, \quad
\ip{U\phi}{U\psi} = \sigma(\ip{\phi}{\psi})  \]
where $\sigma$ is the identity if $U$ is unitary, and complex conjugation if $U$ is antiunitary.
Note that semiunitaries preserve norm, so if $U$ and $V$ are semiunitaries and $U = \lambda V$, then $| \lambda | = 1$.

\begin{theorem}
\label{mainth}
Let $\HH$, $\KK$ be Hilbert spaces of dimension greater than 2. Consider a Chu morphism
\[ (f_*, f^*) :   (\PP(\HH), \LL(\HH), \beH) \rarr (\PP(\KK), \LL(\KK), \beK) . \]
where $f_*$ is injective. Then there is a semiunitary $U : \HH \rarr \KK$ such that $f_* = \PP(U)$. $U$ is unique up to a phase.
\end{theorem}
\begin{proof}
By the proviso on injectivity, we can apply Proposition~\ref{orthpresprop}. By this and Proposition~\ref{projmapprop}, together with the proviso on dimension, we can apply Wigner's Theorem~\ref{Wigth}. Since the semilinear map in Wigner's Theorem has an associated automorphism, we can apply Proposition \ref{surjprop}.
\end{proof}

\subsection{The Representation Theorem}

We now turn to the big picture. We define a category $\SymmH$ as follows:
\begin{itemize}
\item The objects are Hilbert spaces of dimension $> 2$.
\item Morphisms $U : \HH \rarr \KK$ are semiunitary (\ie unitary or antiunitary) maps.
\item Semiunitaries compose as explained more generally for semilinear maps in the previous subsection. Since complex conjugation is an involution, semiunitaries compose according to the rule of signs: two antiunitaries or two unitaries compose to form a unitary, while a unitary and an antiunitary compose to form an antiunitary.
\end{itemize}
This category is a groupoid, \ie every arrow is an isomorphism.

The semiunitaries are the \emph{physically significant symmetries of Hilbert space} from the point of view of Quantum Mechanics. The usual dynamics according to the Schr\"odinger equation is given by a continuous one-parameter group $\{ U(t) \}$ of these symmetries; the requirement of continuity forces the $U(t)$ to be unitaries.\endnote{Indeed, the Schr\"odinger equation can actually be recovered from this group via Stone's Theorem \cite{Sim76}.} However, some important physical symmetries are represented by antiunitaries, e.g. \emph{time reversal} and \emph{charge conjugation}.

By the results of the previous subsection, Chu morphisms essentially force us to consider the symmetries on Hilbert space. As pointed out there, linear maps which can be represented as Chu morphisms in the biextensional category must be injective; and if $g : \HH \rarr \KK$ is an injective linear or antilinear map, then $\PP(g)$ is injective. Our results then show that if $g$ can be represented as a Chu morphism, it must in fact be semiunitary.
This delineation of the physically significant symmetries by the logic of Chu morphisms should be seen as a strong point in favour of this representation by Chu spaces.

We define a functor $R : \SymmH \rarr \eC_{[0, 1]}$:
\[ R : U : \HH \rarr \KK \;\;  \longmapsto \;\; (\Uo, U^{-1}) : (\Ho, \LL(\HH), \eH) \rarr (\Ko, \LL(\KK), \eK) \]
where $\Uo$ is the restriction of $U$ to $\Ho$.

As noted in Proposition~\ref{bcollprop}, the inclusion $\bC_{[0, 1]} \rinc \eC_{[0, 1]}$ has a left adjoint, which we name $Q$. By Proposition~\ref{extprop}, this can be defined on the image of $R$ as follows:
\[ Q : (\Ho, \LL(\HH), \eH) \mapsto (\PP \HH, \LL(\HH), \beH) \]
and for $(\Uo, U^{-1}) : (\Ho, \LL(\HH), \eH) \rarr (\Ko, \LL(\KK), \eK)$,
\[ Q : (\Uo, U^{-1}) \; \longmapsto \; (\PP U, U^{-1}) . \]

We write $\emC$, $\bmC$ for the subcategories of $\eC_{[0, 1]}$ and $\bC_{[0, 1]}$ obtained by restricting to Chu morphisms $f$ for which $f_*$ is injective. The functors $R$ and $Q$ factor through these subcategories.

\begin{proposition}
$R : \SymmH \rarr \emC$ and $Q : \emC \rarr \bmC$ are functors. $R$ is faithful but not full; $Q$ is full but not faithful.
\end{proposition}
\begin{proof}
We  verify that if $U : \HH \rarr \KK$ is semiunitary, $RU$ is a well-defined morphism in $\emC$.
Firstly, we verify the Chu morphism condition. Since $U$ is semiunitary, for $\psi \in \Ho$ and $S \in \LL(\KK)$:
\[ P_{S} (U \psi) = U(P_{U^{-1}(S)} \psi) . \]
Indeed, if $U$ is unitary, let $\{ e_i \}$ be an orthonormal basis for $S$. Then $\{ U^{-1}e_i \}$ is an orthonormal basis for $U^{-1}S$. Now
\[ \begin{array}{rcl}
U(P_{U^{-1}(S)} \psi) & = & U(\sum_i \ip{\psi}{U^{-1}e_i}U^{-1}e_i) \\
& = & \sum_i \ip{\psi}{U^{-1}e_i}e_i \\
& = & \sum_i \ip{U\psi}{e_i}e_i \\
& = & P_S U \psi
\end{array}
\]
where the third equation holds because $U^{-1} = U^{\dagger}$. A similar calculation holds if $U$ is antiunitary. In this case, the inner product is commuted when we apply conjugate linearity in the second equation, and commuted back in the third, since for an antiunitary we have 
\[ \ip{U^{-1}e_i}{\psi} = \ip{U^{-1}e_i}{U^{-1}U\psi} = \ip{U\psi}{e_i} , \]
leading to the same result.

Moreover, $U$ preserves norms, so $\| U\psi \| = \| \psi \|$.
Now
\[ \begin{array}{rcl}
\ip{P_{S} U \psi}{P_{S} U \psi} & = & \ip{U (P_{U^{-1}(S)} \psi)}{U (P_{U^{-1}(S)} \psi)} \\
& = & \ip{P_{U^{-1}(S)} \psi}{P_{U^{-1}(S)} \psi} .
\end{array}
\]
Hence $\eH(\psi, U^{-1}(S)) = \eK(U\psi, S)$, so $(\Uo, U^{-1})$ is a Chu morphism.
Finally, $U$ is bijective, so $\Uo$ is injective.
\end{proof}

We can analyze the non-fullness of $R$ more precisely as follows.
\begin{proposition}
Let $(\Uo, U^{-1}) : (\Ho, \LL(\HH), \eH) \rarr (\Ko, \LL(\KK), \eK)$ be a Chu morphism in the image of $R$.
Given an arbitrary function $f : \HH \rarr \Complex \setminus \{ 0 \}$, define $fU : \Ho \rarr \Ko$ by:
\[ fU(\psi) = f(\psi)U(\psi) .\]
Then $(fU, U^{-1}) \sim (\Uo, U^{-1})$. Moreover, the $\sim$-equivalence class of $U$ is exactly the set of functions of this form.
\end{proposition}
Thus before biextensional collapse, Chu morphisms can introduce arbitrary scalar factors pointwise.
Once we move to the biextensional category, we know by Theorem~\ref{mainth} that our representation will be full, and essentially faithful --- up to a global phase. This points to the need for a projective version of the symmetry groupoid.

The mathematical object underlying phases is the \emph{circle group} $\UG$:
\[ \UG = \{ \lambda \in \Complex \mid |\lambda | = 1 \} = \{ e^{i\theta} \mid \theta \in \Real \} \]
Since $\lambda$ has modulus 1 if and only if $\lambda\bar{\lambda} = 1$, $\UG$ is the unitary group on the one-dimensional Hilbert space.

The circle group acts on the symmetry groupoid $\SymmH$ by scalar multiplication. For Hilbert spaces $\HH$, $\KK$ we can define
\[ \UG \times \SymmH(\HH, \KK) \rarr \SymmH(\HH, \KK) :: (\lambda, U) \mapsto \lambda U . \]
Moreover, this is a category action, meaning that
\[ (\lambda U) \circ V = U \circ (\lambda V) = \lambda (U \circ V). \]
It follows that we can form a quotient category (in fact again a groupoid) with the same objects as $\SymmH$, and in which the morphisms will be the orbits of this group action: 
\[ U \sim V \Iff \exists \lambda \in \UG. \, U = \lambda V . \] 
We call the resulting category $\PSymm$, the \emph{projective quantum symmetry groupoid}. It is a natural generalization of the standard notion of the \emph{projective unitary group} on Hilbert space.
There is a quotient functor $P : \SymmH \rarr \PSymm$, and by virtue of Theorem~\ref{mainth}, we can factor $Q \circ R$ through this quotient to obtain a functor $\PP R : \PSymm \rarr \bmC$.

The situation can be summarized by the following diagram:

\begin{diagram}[heads=vee,width=5em]
\SymmH & \rmon^{R} & \emC \\
\depi<{P} & & \depi>{Q} \\
\PSymm & \rrep_{\PP R} & \bmC
\end{diagram}

\begin{theorem}
\label{repth}
The functor $\PP R : \PSymm \rarr \bmC$ is a representation.
\end{theorem}
\begin{proof}
This follows from Theorem~\ref{mainth}. To see that $\PP R$ is essentially injective on objects, we can use the representation theorems of Piron and Sol\`er \cite{SvS07}, which tell us that we can reconstruct $\HH$ as a Hilbert space from $\LL(\HH)$. This reconstruction will give us a Hilbert space $\HH'$ such that $\LL(\HH) \cong \LL(\HH')$, and $\PP(\HH) \cong \PP(\HH')$. We can apply Wigner's theorem to this isomorphism to obtain a semiunitary $U : \HH \cong \HH'$ from which we can recover the Hilbert space structure on $\HH$. This means that we have recovered $\HH$ uniquely to within the coset of $\id_{\HH}$ in $\PSymm$.
\end{proof}

\section{Reducing The Value Set}
\label{valsect}

We now return to the issue of whether it is necessary to use the full unit interval as the value set for our Chu spaces.

We begin with some generalities.
Given a function $v : K \rarr L$, we define a functor $\Fv : \Chu_{K} \rarr \Chu_{L}$:
\[ \Fv : (X, A, e) \mapsto (X, A, v \circ e) \]
and $\Fv f = f$ for Chu morphisms $f$.

\begin{proposition}
\label{valmapprop}
$\Fv$ is a faithful functor. If $v$ is injective, it is full.
\end{proposition}
\begin{proof}
This is easily verified. The Chu morphism condition is preserved by composing with any function on values, while $\Fv$ is evidently faithful. For fullness, note that the only values in $L$ relevant to whether a pair of functions
\[ (f, g) : (X, A, v \circ e) \rarr (X', A', v \circ e') \]
satisfies the Chu morphism condition are those in the ranges of $v \circ e$ and $v \circ e'$, which if $v$ is injective are in bijection with those in the ranges of $e$ and $e'$.
\end{proof}

We can now state the question we wish to pose more precisely:
\begin{quotation}
Is there a mapping $v : [0, 1] \rarr K$ from the unit interval to some finite set $K$ such that the restriction of the functor $\Fv$ to the image of $\PP R$ is full, and thus the composition
\[ \Fv \circ \PP R : \PSymm \rarr \bmC_{K} \]
is a representation?
\end{quotation}
There is no \emph{general} reason to suppose that this is possible; in fact, we shall show that it is, although not quite in the obvious fashion.

We shall write $\nn = \{ 0, \ldots , n-1 \}$ for the finite ordinals.
The most popular choice of value set for Chu spaces, by far, has been $\Two$, and indeed many interesting categories can be strictly (and even concretely) represented in $\Chu_{\Two}$ \cite{DBLP:conf/lics/Pratt95}.
This makes the following question natural:

\begin{question}
\label{nottwoprop}
Is there a function $v : [0, 1] \rarr \Two$ such that $\Fv \circ \PP R$ is full and faithful?
\end{question}

What we can show is that the most plausible candidates for such functions, yielding the two canonical forms of \emph{possibilistic semantics} as a coarse-graining of probabilistic semantics, both in fact \emph{fail}.

Note that any function $v : [0, 1] \rarr \{ 0, 1 \}$ must lose information either on  $0$ or on $1$ -- or both.
In this sense, the two `sharpest' mappings\endnote{We consider only functions which fix 0 and 1, to exclude irrelevant permutations and the trivial case of constant maps.} will be:
\[  v_0 : 0 \mapsto 0, (0, 1] \mapsto 1  \quad \quad  v_1 : [0, 1) \mapsto 0, 1 \mapsto 1 . \]
These are the two canonical reductions of probabilistic to possibilistic information: the first maps `definitely not' to `no', and anything else to `yes', which is to be read as `possibly yes'; the second maps `definitely yes' to `yes', and anything else to `no', to be read as `possibly no'.
Note that, under the first of these, Lemma~\ref{memblemm} will no longer hold, while under the second, Lemma~\ref{orthlemm} will fail. 

\begin{proposition}
\label{possibprop}
For neither $v = v_0$ nor $v = v_1$ is $F_v \circ \PP R$ full.
\end{proposition}
\begin{proof}
Let $\HH$ be a Hilbert space with $2 < \dim \HH < \infty$, and let $(g, \sigma)$ be any semilinear automorphism of $\HH$, where $\sigma$ can be any automorphism of the complex field.\endnote{We can extend the argument to infinite-dimensional Hilbert space by requiring $g$ to be continuous.} For each of the above two mappings of the unit interval to $\Two$, we shall construct a $\Chu_{\Two}$ endomorphism  $f : F_v \circ \PP R(\HH) \rarr F_v \circ \PP R(\HH)$ with $f_* = \PP(g)$. This will show the non-fullness of $\Fv$.

\textbf{Case 1} Here we consider the mapping $v_1$ which sends $[0, 1)$ to 0 and fixes 1. In this case:
\[ \beH(\ray{\psi}, S) = 1 \IFF \psi \in S \]
and hence the Chu morphism condition on $(f_*, f^*)$, where $f_* = \PP(g)$,  is:
\[ \psi \in f^*(S) \IFF g(\psi) \in S . \]
Taking $f^* = g^{-1}$ obviously fulfills this condition. Note that, since $g$ is a semilinear automorphism, and $\HH$ is finite-dimensional, $g^{-1} : \LL(\HH) \rarr \LL(\HH)$ is well-defined.

\textbf{Case 2} Now consider the mapping $v_0$ keeping 0 fixed and sending $(0, 1]$ to 1. In this case:
\[ \beH(\ray{\psi}, S) = 0 \IFF \psi \orth S \]
and hence the Chu morphism condition on $(f_*, f^*)$, where $f_* = \PP(g)$,  is:
\[ \psi \orth f^*(S) \IFF g(\psi) \orth S . \]
We define $f^*(S) = g^{-1}(S^{\bot})^{\bot}$. Note that $f^* : \LL(\HH) \rarr \LL(\HH)$ is well defined, and also that $g^{-1}(S^{\bot})$ is a subspace of $\HH$; hence 
$g^{-1}(S^{\bot})^{\bot\bot} = g^{-1}(S^{\bot})$.  Now:
\[ \begin{array}{rcl}
\beH(\ray{\psi}, f^*S) = 0 & \IFF & \psi \orth f^*S \\
& \IFF & \psi \in g^{-1}(S^{\bot})^{\bot\bot}  = g^{-1}(S^{\bot}) \\
& \IFF & g(\psi) \in S^{\bot} \\
& \IFF & g(\psi) \orth S \\
& \IFF & \beH(f_*(\ray{\psi}), S) = 0 
\end{array}
\]
and hence $(f_*, f^*)$ is a Chu morphism as required. 
\end{proof}

However, this negative result immediately suggests a remedy: \emph{to keep the interpretations of \textbf{both} 0 \textbf{and} 1 sharp}. We can do this with three values! Namely, we define $v : [0, 1] \rarr \Three$ by
\[ 0 \mapsto 0, \quad (0, 1) \mapsto 2, \quad 1 \mapsto 1 \]
Thus we lose information only on the probabilities strictly between 0 and 1, which are lumped together as `maybe' --- represented here, by arbitrary convention, by 2.

Why is this adequate? Given a vector $\psi$ and a subspace $S$, we can write $\psi$ uniquely as $\theta + \chi$, where $\theta \in S$ and $\chi \in S^{\bot}$. For non-zero $\psi$, there are only three possibilities: $\theta = \zv$ and $\chi \neq \zv$, which yields $\eH(\phi, S) = 0$ by Lemma~\ref{orthlemm}; $\theta \neq \zv$ and $\chi = \zv$ which yields $\eH(\phi, S) = 1$ by Lemma~\ref{memblemm}; and $\theta \neq \zv$ and $\chi \neq \zv$, which yields $\eH(\psi, S) \in (0, 1)$ by these Lemmas again, and hence $v \circ \eH(\psi, S) = 2$. \emph{These are the only case discriminations which are used in our results leading to the Representation Theorem~\ref{repth}}. Hence we have:

\begin{theorem}
\label{threeth}
The functor $\Fv \circ \PP R : \PSymm \rarr \bmC_{\Three}$ is a representation.
\end{theorem}

We note that $\Chu_{\Three}$ has found some uses in concurrency and verification \cite{DBLP:journals/mscs/Pratt03,DBLP:conf/cdes/Ivanov08}, under a temporal interpretation: the three values are read as `before', `during' and `after', whereas in our setting the three values represent `definitely yes', `definitely no' and `maybe'.

Theorem~\ref{threeth} may suggest some interesting uses for 3-valued `local logics' in the sense of Jon Barwise \cite{Barwise97}.

\section{Discussion}

We should understand Chu spaces as providing a very general (and, we might reasonably say, rather simple)  `logic of systems or structures'.
Indeed, they have been proposed by Barwise and Seligman as the vehicle for a general logic of `distributed systems' and information flow \cite{Barwise97}. This logic of Chu spaces was in no way biassed in its conception  towards the description of quantum mechanics or any other kind of physical system.
Just for this reason, it is interesting to see how much of quantum-mechanical structure and concepts can be absorbed and essentially \emph{determined} by this more general systems logic.

It might be argued that our representation of quantum systems as Chu spaces has already specified the essential ingredients of the quantum structure `by hand'. The conceptual significance of our technical results is precisely to show that there is a non-trivial `capturing' of quantum structure by the general notions of Chu spaces:
\begin{itemize}
\item Firstly, Proposition~\ref{extprop} shows that the general Chu space notion of biextensionality subsumes the standard identification of quantum states with rays in Hilbert space.
This is scarcely surprising, but it is a first sign of the proper alignment of concepts.

\item The main technical result of the present paper is the Representation Theorem~\ref{repth}.
It is worth spelling out the content of this in more elementary terms.
Once we have represented our quantum systems as biextensional Chu spaces $(\PP(\HH), \LL(\HH), \eH)$, all we have, from the viewpoint `inside' the category $\Chu_{[0, 1]}$, is a pair of sets and an evaluation function, with all information about their provenance lost. A Chu morphism 
\[ (f_*, f^*) : (\PP(\HH), \LL(\HH), \eH) \rarr (\PP(\KK), \LL(\HH), \eK) \]
is given by \emph{any} pair of set-theoretic functions $(f_*, f^*)$ satisfying the Chu morphism condition:
\[ \beH(\ray{\psi}, f^*(S)) = \beK(f_*(\ray{\psi}), S) . \]
The Representation Theorem says that \emph{the logic of this Chu morphism condition is strong enough to guarantee that any such pair of functions must arise from a unitary or antiunitary map $U : \HH \rarr \KK$ on the underlying Hilbert spaces}, with the sole proviso of injectivity of $f_*$.\endnote{The injectivity assumption on $f_*$ is annoying. It remains unclear if it necessary.} Moreover, $U$ is uniquely determined by $f_*$ up to a phase factor. Of course, we are using one of the `big guns' of the subject, Wigner's Theorem, to establish this result. It is worth noting, though, that there is some distance to travel between the Chu morphism condition and the hypotheses of Wigner's Theorem; and there are surprises along the way, most notably Proposition~\ref{surjprop}, which \emph{derives} surjectivity from the Chu morphism condition --- whereas it must invariably be added as a hypothesis to the many versions of Wigner's Theorem.\endnote{One of the journal referees remarked that surjectivity is not taken as a hypothesis in the version of Wigner's Theorem due to Wright \cite{Wright}. Wright's Corollary~1 does obtain surjectivity from the assumption that the function $f : \PP(\HH) \rarr \PP(\KK)$ on rays can be extended to a \emph{projection-valued state}, \ie a map  $\LL(\HH) \rarr \LL(\KK)$ which preserves orthogonal joins and the top element. These are strong assumptions, from which surjectivity follows immediately, as Wright observes. By contrast, the injectivity of $f$ and the Chu morphism condition, with respect to an a priori otherwise unspecified map $\LL(\KK) \rarr \LL(\HH)$, are much weaker assumptions, which do not `build in' surjectivity in any obvious fashion.}

\item The results on reduction to finite value sets are also intriguing. Not only is the bare Chu condition on morphisms sufficient to whittle them down to the semiunitaries, this is even the case when the discriminations on which the condition is based  are reduced to three values. The general case for two values remains open, but we have shown that the two standard possibilistic reductions both \emph{fail to preserve fullness}. A negative answer for two-valued semantics in general would suggest an unexpected r\^ole for three-valued logic in the foundations of Quantum Mechanics.

\end{itemize}

\paragraph{Where Next?}

Of course, the developments described in the present paper are only a first step.
We shall briefly discuss some of the natural continuations of these ideas, several of which are already in progress.

\begin{itemize}
\item There are some interesting and surprising connections  between Chu spaces and another important paradigm for categorical systems modelling, namely \emph{coalgebra} \cite{DBLP:journals/tcs/Rutten00}. These connections,  which seem not to have been explored previously, arise both at the general level, and also with specific reference to the representation of physical systems. They are described in a  sequel to the present paper \cite{CoChu}, which lifts the results of the present paper to a coalgebraic setting. The bivariant nature of Chu spaces is reflected in  a novel fibred form of coalgebra, in which contravariance is represented as \emph{indexing}.
\item A natural next step as regards physical modelling is to consider \emph{mixed states}. There is a general construction on Chu spaces which allows mixed states to be studied in a uniform fashion, applicable to both classical and quantum systems. This will be described in a forthcoming sequel to the present paper.
\item There are intriguing connections between our approach, and the work of the `Geneva School' of Jauch and Piron \cite{Jauch,Piron}, particularly  \cite{FMP}. We plan to explore these in a joint paper with Bob Coecke, Isar Stubbe and Frank Valckenborgh.
\item It is also of interest to consider \emph{universal} Chu spaces; single systems in which all Chu spaces of a given class can be embedded, and which therefore provide a single model for a large class of systems. We may additionally ask for such systems to be \emph{homogeneous}, which means that they exhibit a maximum degree of symmetry; such universal, homogeneous spaces are unique up to isomorphism. Universal homogeneous Chu spaces have been constructed for \emph{bifinite Chu spaces} in recent work by Manfred Droste and Guo-Qiang Zhang \cite{DBLP:conf/calco/DrosteZ07}. That context is too limited for our purposes here. It remains to be seen if universal homogeneous models can be constructed for larger subcategories of Chu spaces, encompassing those involved in our representation results.

\item The relation of the rich logical and type-theoretic aspects of Chu spaces to quantum and other physical systems should also be investigated.
\end{itemize}

\theendnotes

\bibliographystyle{klunamed}

\bibliography{bibfile}

\end{article}
\end{document}